\documentclass[amssymb,prl,twocolumn,superscriptaddress,showpacs]{revtex4}
\usepackage{graphicx}
\usepackage{dcolumn}
\usepackage{epsfig}
\usepackage{bm}
\usepackage{color}
\usepackage[colorlinks]{hyperref}
\usepackage{verbatim}
\usepackage[sort&compress]{natbib}
\begin{document}


\title{\bf Multiple photoexcitation of  two-dimensional electron systems:
bichromatic magnetoresistance oscillations revisited }

\author{Jes\'us I\~narrea}
\affiliation {Escuela Polit\'ecnica Superior,Universidad Carlos
III,Leganes,Madrid,28911,Spain} \affiliation{Unidad Asociada al
Instituto de Ciencia de Materiales, CSIC,
Cantoblanco,Madrid,28049,Spain.}

%
%
%
%
\date{\today}
\begin{abstract}
We analyze theoretically magnetoresistance of high mobility
two-dimensional electron systems being illuminated by multiple
radiation sources. In particular, we study the influence on the
striking effect of microwave-induced resistance oscillations. We
consider moderate radiation intensities without reaching the zero
resistance states regime. We use the model of radiation-driven
Larmor orbits extended to several light sources. First, we study the
case of two different radiations polarized in the same direction
with different or equal frequencies. For both cases we find a regime
of superposition or interference of harmonic motions. When the
frequencies are different, we obtain a modulated magnetoresistance
response with pulses and beats. On the other hand, when the
frequencies are the same, we find that the final result will  depend
on the phase difference between both radiation fields going from an
enhanced response to a total collapse of oscillations, reaching an
outcome similar to darkness. Finally, we consider a multiple
photoexcitation case (three different frquencies) where we propose
the two-dimensional electron system as a potential nanoantenna
device for microwaves.

\end{abstract}
%
%
\maketitle
\section{I. introduction}
Photo-excited transport in two-dimensional electron system (2DES) is
currently a fundamental topic from experimental and theoretical
perspectives\cite{ina1}. This interest comes not only from the basic
explanation of a physical effect but also from the potential
application for the development of future technological devices.
When a Hall bar (a 2DES with a uniform and perpendicular magnetic
field ($B$)) is irradiated with microwaves, different effects can be
observed. Among them special attention deserves the recently
discovered, microwave-induced (MW) resistance oscillations (MIRO)
and zero resistance states (ZRS)
\cite{mani1,zudov1,studenikin1}. Those effects, that show up
at low $B$ in high mobility samples, caused a big impact in the
condensed matter community, mainly because they were obtained
without quantization in the Hall resistance. Different theories have
been proposed to explain the origin and physical consequences of
these striking effects
\cite{ina2,girvin,dietel,lei,ryzhii,rivera,shi,mirlin} but the
physical origin is still being questioned and, in spite of the
progress, there remain many aspects that could be better understood.

Thus, to unveil the physics behind them, a great effort has been
made specially from the experimental side adding new features and
different probes to the basic experimental setup\cite{mani2,
mani3,willett,mani4,smet,yuan,stone,doro,hatke,mani5}. The obtained
experimental results are real challenges for the available
theories. Thus, as an example,
it has been recently published experimental results on the
dependence of the oscillations with radiation power\cite{mani6}
where a very solid result has been obtained in terms of a sublinear relation, similar
to a square root. This result has been also recently confirmed by
other experiments on two-dimensional bilayer systems\cite{wiedmann}.
Yet, some theories predicted a linear dependence\cite{dmitriev} between
MIRO and radiation power.

One of the most interesting setups that has been carried out,
consists in illuminating the sample with two different light
sources\cite{mani7}. Some of the experimental results were realized
at high radiation intensities making the MW-response to evolve into
the  zero resistance states regime\cite{zudov2}. The unexpected
obtained results consisted in a transformed magnetoresistance
($R_{xx}$) profile with new features including  different peak
positions and intensities and new zero resistance state regions. At
first they were explained in terms of a theory of current domains
formation\cite{mirlin,kunold}. Other theories offered an alternative
explanation based in the superposition of two radiation-driven
motions acting on the center of the Larmor orbits\cite{ina3}. Yet, a
regime of moderate radiation intensities without reaching zero
resistance states has not been yet sufficiently studied in
experiments. We think that, if such an experiment were carried out,
their results, about how the resistance profile is transformed,
would shed some light on the influence of adding extra light sources
on the magnetoresistance and, in the end, in the origin of MIRO. Of
course the obtained experimental results would mean a real challenge
for the existent theoretical models. Thus, a comparison of
experiment with theory could help to identify the importance of the
invoked-mechanisms in these theories.
On the other hand, theories not only have to reasonably explain some
experimental results. A solid theoretical model has to offer also
definite predictions to be obtained if some parameters or variables
are changed in a certain experimental set up. These predictions
serve as orientation to experimentalists. And, if the theoretical
predictions are confirmed, they also help to identify and interpret
which mechanism is behind some definite physical effect. This is the
main goal of the present article, where we offer some theoretical
predictions about the changes to be produced in MIRO when a 2DES is
subjected to multiple radiations. Thus, in this article, we
theoretically study magnetoresistance of a Hall bar being
illuminated simultaneously by several radiation sources. We apply
the model developed by the authors, the MW-driven Larmor
orbits\cite{ina2,ina3}, which is extended to a multiexcitation
situation with moderate radiation intensities. According to this
theory\cite{ina2,ina3} when a Hall bar is illuminated, the
corresponding orbit centers of the Landau states (electronic
oscillators) perform a classical trajectory consisting in a harmonic
 motion in the direction of the current. Thus, the whole 2DES moves
periodically at the MW frequency. Our work is just focussed on how
this response is going to be altered by the presence of extra
radiation sources and eventually how this will be reflected in the
magnetoresistance oscillations. We predict that in the presence of
several radiation fields, the final motion of the center of the
Larmor orbits will consist in the superposition of several harmonic
motions, giving rise to interference effects. Accordingly, different
response will be obtained depending on the relative frequencies,
phase difference, radiation intensities, etc. We begin by
considering only two sources of radiation that can have different or
equal frequencies. In the first case we obtain a modulated $R_{xx}$
response with pulses and beats. In the second case, the phase
difference between the two light sources plays a key role. Thus,
depending on its value we can achieve from an enhanced response
(constructive interference) to a total collapse of the oscillations
(destructive interference) with a result similar to darkness. This
is a striking result that has not been obtained yet in experiments.
Finally, we consider a multiple photoexcitation (three light
sources) case with different frequencies where we propose the
two-dimensional electron system as a multifrequency radiation sensor
or nanoantenna device for the microwave range of radiation.

A potential experiment with phase difference could be envisaged as follows.
The idea would be  to start with one microwave source. Then split the microwave beam into two parts using a "splitter", one that splits an input signal into two equal phase output signals. Then insert a phase shifter in
the path of both beams. The reason to put it in the path of both beams
rather than just one is because the phase shifter, even when it is not
shifting phase, will introduce some power loss, and we want matched power in
both beams.
The phase shifter can be used to introduce an arbitrary phase $(0^{0}<\theta <
180^{0} ) $ shift in either beam by turning a knob. Next, we can take the
two beams separately to the sample and illuminate the sample with them.

Microwave-driven Larmor orbits model describe MIRO as a borderline effect between Classical Mechanics and Quantum Mechanics. Thus, this model assigns an essential classical feature to MIRO, such as the classical trajectory that the electron orbits center guide performs driven by radiation. The rest of the model is basically based in Quantum Mechanics.
The other models presented in the literature can be classified in either "displacement models"\cite{girvin}
or "inelastic models"\cite{dmitriev}. These two theories are fully based in Quantum Mechanics.
The results predicted in this paper are directly connected to the classical trajectory of the electronic orbits. Thus, these expected effects, that depend on equal or different frequencies and in the relative values of phase differences, are totally classical effects.
Therefore if predictions are correct, neither the displacement model nor the inelastic mechanism would be able to
fully  explain the physics beneath them.

\begin{figure} \centering\epsfxsize=3.5in
\epsfysize=5.0in \epsffile{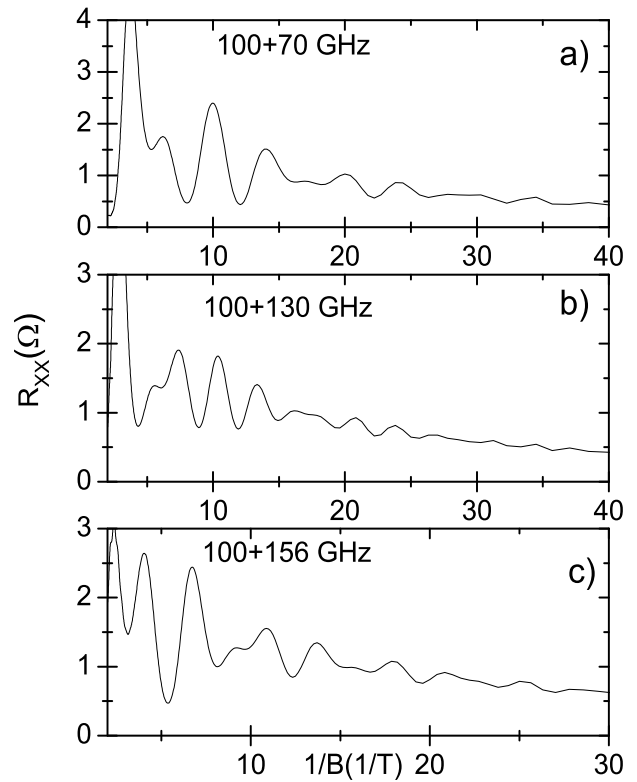} \caption{
Calculated longitudinal magnetoresistance versus
the inverse of the magnetic field for two light sources having different frequencies.
These frequencies, from a) to c) panels, are: $100+70$ GHz,
$100+130$ GHz and $100+156$ GHz. We have used large
MW frequencies to have a sufficient number of oscillations
to observe oscillations beats.
The peak
separation is going to be the same making beats
more clearly seen in the three panels, although the peak intensity is smeared out
for higher values of $1/B$.  
}
\end{figure}

\section{II. Theoretical model}
The {\it MW driven Larmor orbits model}, was developed to explain
the $R_{xx}$ response of an irradiated 2DEG at low $B$. We obtained
the exact solution of the corresponding electronic wave function
when the 2DES is being illuminated by
radiation\cite{ina2,ina3,ina4,kerner,park}:
\begin{eqnarray}
\Psi_{N}(x,t)\propto\phi_{n}(x-X-x_{cl}(t),t)
\end{eqnarray}
, where $\phi_{n}$ is the solution for the
Schr\"{o}dinger equation of the unforced quantum harmonic
oscillator, $X$ is the center of the orbit for the electron
motion. $x_{cl}(t)$ is the classical solution of a forced  harmonic
oscillator\cite{ina2,ina3,kerner,park}. When one considers that the system
is being driven by two different time dependent forces (two light sources)
$x_{cl}(t)$ has now the expression:
\begin{eqnarray}
x_{cl}(t)&=&x_{1}(t)+x_{2}(t)= \nonumber\\
&=&\frac{e
E_{1}}{m^{*}\sqrt{(w_{c}^{2}-w_{1}^{2})^{2}+\gamma^{4}}}\cos w_{1}t\nonumber\\
&+&
\frac{e E_{2}}{m^{*}\sqrt{(w_{c}^{2}-w_{2}^{2})^{2}+\gamma^{4}}}\cos
w_{2}t =\\
&=&A_{1}\cos w_{1}t +  A_{2}\cos w_{2}t
\end{eqnarray}

where $e$ is the electron charge, $\gamma$ is a
phenomenologically-introduced damping factor for the electronic
interaction with the lattice ions emitting acoustic phonons and $w_{c}$ is the
cyclotron frequency. $E_{1}$ and $E_{2}$ are the  electric
fields amplitudes of the two radiation sources.  Then, the obtained wave
function is the same as the standard harmonic oscillator where the
center is displaced by $x_{cl}(t)$. Thus, the orbit centers are not
fixed, but they oscillate harmonically at the radiation field
frequency $w$, if we only had one radiation field. In the case of two
radiation fields, the classical trajectory of the orbit center would be
the result of the superposition of two harmonic motions producing
different types of interference effects that will be reflected in the final $R_{xx}$
response.

This $radiation-driven$ behavior will affect dramatically the
charged impurity scattering and eventually the conductivity.
Therefore,
 we introduce the scattering suffered by the electrons due to
charged impurities randomly distributed in the sample.
 If the scattering is weak, we
can apply time dependent first order perturbation theory. Thus, first we
calculate the impurity scattering rate $W_{N,M}$
between two $oscillating$ Landau states $\Psi_{N}$, and
$\Psi_{M}$\cite{ina2,ina3,riddley,davies,ando}.
Next we find the average effective distance advanced by the electron
in every scattering jump
that in the case of two MW
sources is given by:
\begin{equation}
 \Delta X^{MW}=\Delta X^{0}+ A_{1}\cos
w_{1}\tau+A_{2}\cos w_{2}\tau
\end{equation}
, where $\Delta X^{0}$ is the effective distance advanced when there
is no MW field present and $\tau=1/W_{N,M}$ is the scattering time.
Finally the longitudinal conductivity $\sigma_{xx}$ is given by:
$\sigma_{xx}\propto \int dE \frac{\Delta X^{MW}}{\tau}$
being $E$
the energy.
To obtain $R_{xx}$ we use
the relation
$R_{xx}=\frac{\sigma_{xx}}{\sigma_{xx}^{2}+\sigma_{xy}^{2}}
\simeq\frac{\sigma_{xx}}{\sigma_{xy}^{2}}$, where
$\sigma_{xy}\simeq\frac{n_{i}e}{B}$ and $\sigma_{xx}\ll\sigma_{xy}$.

\begin{figure}
\centering\epsfxsize=3.5in \epsfysize=5.5in
\epsffile{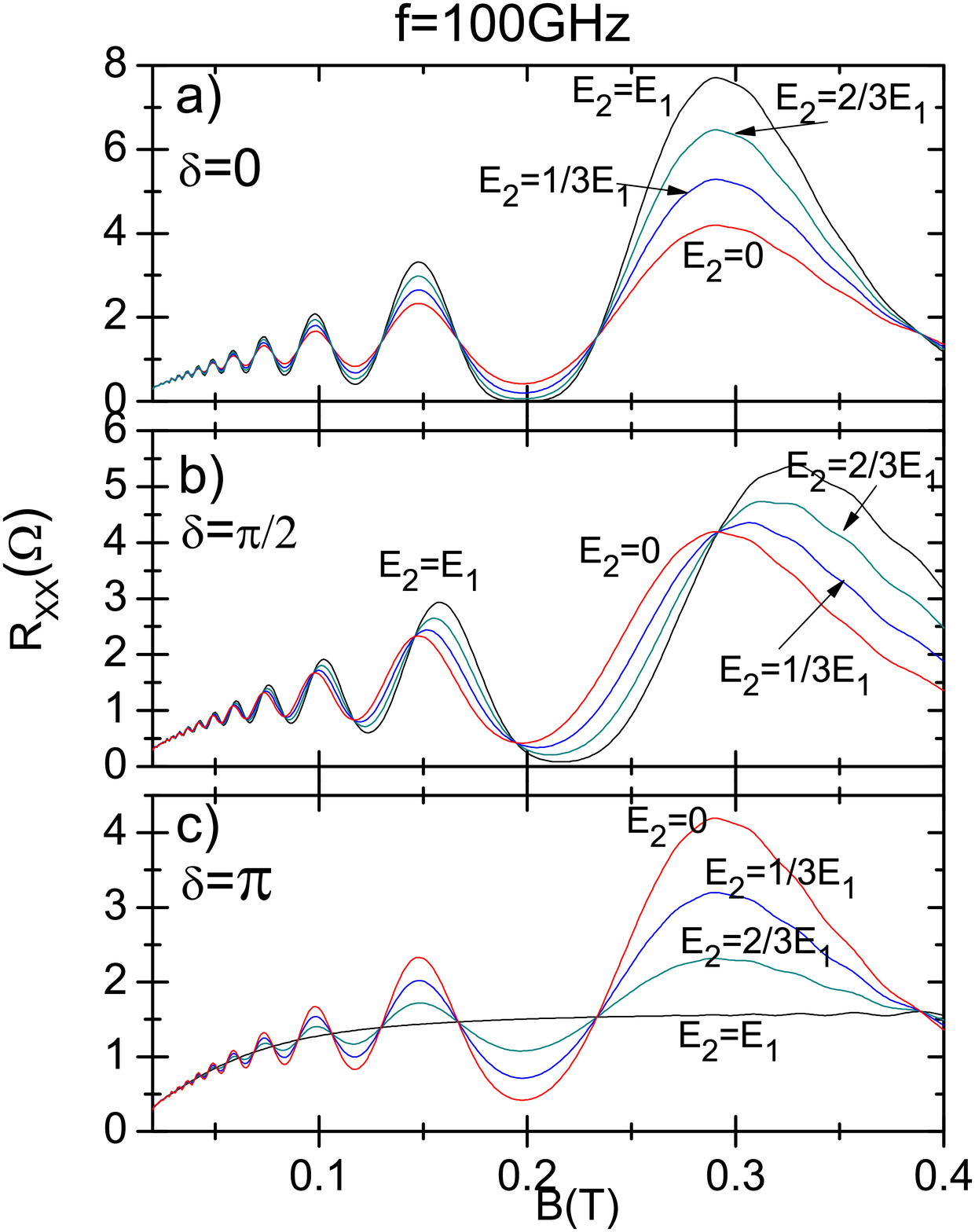} \caption{
Calculated $R_{xx}$ versus magnetic field
for two different light sources illuminating the
sample.
Fig. 2a, the phase difference $\delta=0$,
the interference is constructive giving an enhanced
$R_{xx}$ response. $E_{1}$ and $E_{2}$ represent
the corresponding radiation electric fields.
From $E_{2}=0$  to $E_{2}=E_{1.}$,
MIRO amplitudes increase, reaching ZRS around
$B=0.2T$. Fig. 2b,
$\delta=\pi/2$. As
$E_{2}$ increases, the amplitude of,
MIRO also increases and for $E_{2}=E_{1}$, ZRS are
almost reached at about  $B=0.22T$. On the other hand,
there is also an increasing shift to the right of the whole MIRO
response coming from the phase difference which
shows up in equation (14).  Fig. 2c,
$\delta=\pi$, the interference is destructive and
as the intensity $E_{2}$ increases, MIRO are
progressively smaller till the oscillations totally
collapse. }
\end{figure}

Now we can proceed to consider some interesting cases depending on
the relative frequencies of the light sources, phase difference,
intensities etc. In all cases we are going to find that the center
of the electronic orbits is subject to more than one harmonic
time-dependent force, each trying to move the center in its own
direction, giving rise to interference effects. Eventually this will observed in the
$R_{xx}$ oscillations producing different responses depending
on the relative values (frequency, phase difference, intensity)
of the radiations fields. Thus, if $w_{1}$ is
not very different from $w_{2}$ and the MW fields intensities are
equal, then we can write, $A_{1}\simeq A_{2}=A$ and therefore:
\begin{eqnarray}
x_{cl}(t)&=&A [\cos w_{1}t+ \cos w_{2}t]\nonumber\\
&=& 2A \cos\left [\frac{1}{2}(w_{1}-w_{2})t\right ] \cos \left
[\frac{1}{2}(w_{1}+w_{2})t\right]
\end{eqnarray}
showing that now the oscillatory movement for the Larmor orbits
center presents modulated amplitude with a frequency given by
$\frac{1}{2}(w_{1}-w_{2})$ whereas the main oscillation goes like
$\frac{1}{2}(w_{1}+w_{2})$. This results is reflected in the average
advanced distance by the electron in each scattering jump:
\begin{equation}
\Delta X^{MW}=\Delta X^{0}+2A \cos\left
[\frac{1}{2}(w_{1}-w_{2})\tau\right ] \cos \left
[\frac{1}{2}(w_{1}+w_{2})\tau\right]
\end{equation}
\\

Thus, in this case we will observe $R_{xx}$ oscillations with
pulses and beats.

Another interesting regime takes place when the two radiations fields have the
same frequency: $w_{1}=w_{2}$. As in the latter case we have an interference
effect which now depends on the phase difference ($\delta$) between the two light
sources and that can be tuned to obtain opposite effects in the $R_{xx}$ response. The phase difference
between the radiation fields is
translated into the expression of $x_{cl}$ writing now:
\begin{equation}
x_{cl}(t)=A_{1}\cos wt +  A_{2}\cos (wt+\delta)
\end{equation}

If $\delta=0$ $\Rightarrow$ $x_{cl}(t)=(A_{1}+A_{2})\cos wt $ and the interference
is constructive. Then, the
average distance advanced in every scattering event in the direction of
the current will be
\begin{equation}
\Delta X^{MW}=\Delta X^{0}+(A_{1}+A_{2})\cos w\tau
\end{equation}
Therefore, we
will obtain an enhanced response in $R_{xx}$ regarding
the case of just one light source with the same frequency
and intensity.\\ If $\delta=\pi$ $\Rightarrow$ $x_{cl}(t)=(A_{1}-A_{2})\cos wt $ and the interference
is destructive. Then,
\begin{equation}
\Delta X^{MW}=\Delta X^{0}+(A_{1}-A_{2})\cos w\tau
\end{equation}
 Therefore, we
will obtain a reduced response in $R_{xx}$ regarding
the case of just one light source with the same frequency
and intensity. In the particular case or equal intensities, $A_{1}\simeq A_{2}$, we
will obtain a striking result in the nearly total destruction of MIRO and the
response will be the same as in darkness, $\Delta X^{MW}=\Delta X^{0}$.\\
For an arbitrary value of the phase difference is straightforward to obtain
a simpler  expression for $x_{cl}$ with new amplitude $A$ and phase difference $\alpha$:
\begin{equation}
x_{cl}(t)=A \cos (wt+\alpha)
\end{equation}
where,
\begin{equation}
A=\sqrt{A_{1}^{2}+A_{2}^{2}+2A_{1}A_{2} \cos \delta}
\end{equation}
and,
\begin{equation}
\tan \alpha=\frac{A_{2}\sin \delta}{A_{1}+A_{2}\cos \delta}
\end{equation}
Therefore, for $\delta=\frac{\pi}{2}$, we readily obtain,
\begin{eqnarray}
x_{cl}(t)=\sqrt{A_{1}^{2}+A_{2}^{2}} \cos \left(wt+\arctan \left[\frac{A_{2}}{A_{1}} \right]\right)\\
\Delta X^{MW}=\sqrt{A_{1}^{2}+A_{2}^{2}} \cos \left(w\tau+\arctan\left[\frac{A_{2}}{A_{1}}\right]\right)
\end{eqnarray}
Then, we expect for this case not only a different MIRO amplitude but also
an oscillations shift due to the presence of the phase difference $\alpha$.

The theory can be extended to a higher number of radiation sources to have a multiexcitation
of the sample. Thus, we can write for $n$ sources:
\begin{eqnarray}
x_{cl}(t)=\sum_{i=1}^{n} A_{i} \cos w_{i}t\\
\Delta X^{MW}=\sum_{i=1}^{n} A_{i} \cos w_{i}\tau
\end{eqnarray}
In this case we will have a linear superposition of all
radiations giving rise to a more pronounced interference effects.


\section{Calculated results}
In Fig. 1 we present calculated longitudinal magnetoresistance versus
the inverse of the magnetic field for two light sources having different frequencies.
These frequencies, from a) to c) panels, are: $100+70$ GHz,
$100+130$ GHz and $100+156$ GHz. We clearly observe that
the amplitudes are modulated in the three
pairs of frequencies. We have used large
MW frequencies to have a sufficient number of oscillations
that permit us to observe oscillations beats.
Versus ($1/B$) the peak
separation is going to be the same making beats
more clearly seen in the three panels, although the peak intensity is smeared out
for higher values of $1/B$.  

In Fig. 2 we present calculated $R_{xx}$ versus magnetic field
for two different light sources
 with the same frequency. We observe that depending on
the phase difference ($\delta$) between the two radiation fields
different outcomes are obtained. Thus, in Fig. 2a, $\delta=0$,
the interference is constructive, (see eq. (9)), and we obtained an enhanced
$R_{xx}$ response. In the figure $E_{1}$ and $E_{2}$ represent
the corresponding radiation electric fields.
We represent curves from $E_{2}=0$  to $E_{2}=E_{1.}$,
observing that MIRO amplitudes increase and for
equal electric field intensities, ZRS are reached, around
$B=0.2T$. In Fig. 2b,
$\delta=\pi/2$, we represent the same cases as in 2b. We observe that as
$E_{2}$ increases two effects take place. First, the amplitude of
MIRO increases, as expected from equation (14), and for $E_{2}=E_{1}$, ZRS are
almost reached around $B=0.22T$. Second,
there is also an increasing shift to the right of the whole $R_{xx}$
response coming from the phase difference, $\arctan \left[\frac{A_{2}}{A_{1}} \right]$,  which
shows up in equation (14).  Finally,
in Fig. 2c,
$\delta=\pi$, the interference is destructive (see eq. (9)) and,
 as the intensity $E_{2}$ increases, MIRO are
progressively smaller till the oscillations totally
collapse. Thus, although the sample is illuminated,
the obtained $R_{xx}$ response corresponds to darkness.

As we said above, the theory can be extended to a higher number of
radiation sources. If all radiations have different frequencies we
can define it as {\it multichromatic excitation} of the sample: $\Delta
X^{MW}=\sum_{i=1}^{n} A_{i} \cos w_{i}\tau$. In this case we will
have a linear superposition of all radiations giving rise to more
pronounced interference effects. As an illustrative example we can
consider the case of three radiation waves with different frequencies. The
most interesting case could be of an {\it amplitude modulated radio
signal}. It is well known that the electric field of such a radiation can be mathematically
represented by\cite{comm}:
\begin{equation}
E(t)=B_{1}\cos w_{1}t+\frac{1}{2}B_{2}(\cos w_{2}t+\cos w_{3}t)
\end{equation}
which is the sum of three harmonic contributions of different
frequencies. Therefore, the modulated signal has three harmonic
components, a carrier wave and two sinusoidal waves known
as sidebands whose frequencies are slightly above and below of
the carrier wave frequency. In this particular case these frequencies are related
among them by definite relations: $w_{1}=w_{c}$ which is
the frequency of the carrier wave, then  $w_{c}$ is
known as {\it carrier wave frequency}. The other two
frequencies corresponds to the sidebands: $w_{2}=w_{c}+w_{m}$ and
$w_{2}=w_{c}-w_{m}$ where
$w_{m}$ is known as {\it message wave frequency}. $B_{1}$ and
$B_{2}$ are the corresponding amplitudes. 


According to the above results, we have shown that the
electronic oscillators can be  driven by more than one
electromagnetic waves,
absorbing the corresponding energy from
them\cite{ina5}. The theoretical model of {\it Radiation
driven Larmor orbits} states that the electromagnetic
radiation will translate electric field intensity and
frequency into the motion of the electronic orbits.
In other words, {\it information} transported in the
electromagnetic waves in form of amplitude of
frequency can be transferred in an efficient
way into the 2DES through the center guide of
the electronic orbits. Therefore, we have
demonstrated that a hall bar can perform efficiently
as a multifrequency radiation sensor or nanoantenna
in the MW range of radiation.
Another important issue with interest not
only from basic knowledge but also from a technological
standpoint is the possibility of reemission for
these systems. The topic of reemission has not been
yet addressed by any theory. Yet, the
microwave-driven Larmor orbit model can
do it considering that the radiation-driven
back and forth motion of the electronics orbits
makes them oscillating dipoles. Importantly,
the possibility of harmonic excitation and frequency
multiplication with the corresponding reemission
can be addressed by the present model. These features
have been already predicted by this theory\cite{ina6}
when these systems reach slightly anharmonic regimes.
All these topics constitute the core of a future
work.

\section{Conclusions}
In summary, We have studied $R_{xx}$ of high mobility
2DES being illuminated by multiple
radiation sources. We have essentially studied the influence on the
microwave-induced resistance oscillations in a moderate intensity
radiation regime, which excludes ZRS.
We have applied the model of radiation-driven
Larmor orbits extended to several radiations (multichromatic). First, we study the
case of two different radiations
with different or equal frequencies. For both cases we find a regime
of  interferences of harmonic motions acting on the center guide
of the electronic orbits. When the
frequencies are different, we obtain a modulated magnetoresistance
response with pulses and beats. On the other hand, when the
frequencies are the same, we find that the result will  depend
on the phase difference between both radiation fields going from an
enhanced response to a total collapse. Finally, we consider a multiple
photoexcitation case where we propose
the two-dimensional electron system as a potential nanoantenna
device for microwave radiation.

J.I. is supported by the MCYT (Spain) under grant:
MAT2008-02626/NAN.

\end{document}